\documentclass[prb,superscriptaddress,twocolumn,floatfix,amsmath,amssymb,showpacs]{revtex4-1}

\usepackage{multirow}
\usepackage{epstopdf}
\usepackage{graphicx}
\usepackage{dcolumn}
\usepackage{bm}
\usepackage{color}
\hfuzz=\maxdimen
\begin{document}

\title{Metal-to-metal transition and heavy-electron state in Nd$_4$Ni$_3$O$_{10-\delta}$}

\author{Bai-Zhuo Li}
\affiliation{Department of Physics, Zhejiang Province Key Laboratory of Quantum Technology and Devices, Interdisciplinary Center for Quantum Information, and State Key Lab of Silicon Materials, Zhejiang University, Hangzhou 310027, China}
\author{Cao Wang}
\affiliation{School of Physics \& Optoelectronic Engineering, Shandong University of Technology, Zibo 255000, China}
\author{P. T. Yang}
\affiliation{Beijing National Laboratory for Condensed Matter Physics and Institute of Physics, Chinese Academy of Sciences, Beijing, 100190, China}
\affiliation{School of Physical Sciences, University of Chinese Academy of Sciences, Beijing 100190, China}
\author{J. P. Sun}
\affiliation{Beijing National Laboratory for Condensed Matter Physics and Institute of Physics, Chinese Academy of Sciences, Beijing, 100190, China}
\affiliation{School of Physical Sciences, University of Chinese Academy of Sciences, Beijing 100190, China}

\author{Ya-Bin Liu}
\affiliation{Department of Physics, Zhejiang Province Key Laboratory of Quantum Technology and Devices, Interdisciplinary Center for Quantum Information, and State Key Lab of Silicon Materials, Zhejiang University, Hangzhou 310027, China}
\author{Jifeng Wu}
\affiliation{School of Science, Westlake Institute for Advanced Study, Westlake University, Hangzhou 310064, China}
\author{Zhi Ren}
\affiliation{School of Science, Westlake Institute for Advanced Study, Westlake University, Hangzhou 310064, China}
\author{J. -G. Cheng}
\affiliation{Beijing National Laboratory for Condensed Matter Physics and Institute of Physics, Chinese Academy of Sciences, Beijing, 100190, China}
\affiliation{School of Physical Sciences, University of Chinese Academy of Sciences, Beijing 100190, China}
\affiliation{Songshan Lake Materials Laboratory, Dongguan, Guangdong 523808, China}
\author{Guang-Ming Zhang}
\affiliation{State Key Laboratory of Low-Dimensional Quantum Physics and
Department of Physics, Tsinghua University, Beijing 100084, China}
\author{Guang-Han Cao}
\email[corresponding author: ]{ghcao@zju.edu.cn}
\affiliation{Department of Physics, Zhejiang Province Key Laboratory of Quantum Technology and Devices, Interdisciplinary Center for Quantum Information, and State Key Lab of Silicon Materials, Zhejiang University, Hangzhou 310027, China}
\affiliation{Collaborative Innovation Centre of Advanced Microstructures, Nanjing 210093, China}

\date{\today}

\begin{abstract}
The trilayer nickelate Nd$_4$Ni$_3$O$_{10-\delta}$ ($\delta \approx$ 0.15) was investigated by the measurements of x-ray diffraction, electrical resistivity, magnetic susceptibility, and heat capacity. The crystal structure data suggest a higher Ni valence in the inner perovskite-like layer. At ambient pressure the resistivity shows a jump at 162 K, indicating a metal-to-metal transition (MMT). The MMT is also characterized by a magnetic susceptibility drop, a sharp specific-heat peak, and an isotropic lattice contraction. Below $\sim$ 50 K, a resistivity upturn with a log$T$ dependence shows up, accompanying with a negative thermal expansion.  External hydrostatic pressure suppresses the resistivity jump progressively, coincident with the diminution of the log$T$ behavior. The low-temperature electronic specific-heat coefficient is extracted to be $\sim$ 150 mJ K$^{-2}$ mol-fu$^{-1}$, equivalent to $\sim$ 50 mJ K$^{-2}$ mol-Ni$^{-1}$, indicating an unusual heavy-electron correlated state. The novel heavy-electron state as well as the logarithmic temperature dependence of resistivity is explained in terms of the Ni$^{3+}$ centered Kondo effect in the inner layer of the (NdNiO$_3$)$_3$ trilayers.
\end{abstract}


\maketitle
\section{\label{sec:level1}Introduction}

Perovskite-like nickelates possess similar crystal and electronic structures with those of high-$T_\mathrm{c}$ cuprate superconductors~\cite{43d.nc2017}. It was earlier theoretically expected that high-$T_\mathrm{c}$ superconductivity could be realized in layered nickelates with NiO$_2$ sheets~\cite{Rice.prb1999,Khaliullin.prl2008,PRM2017.Norman}, although the opposing ideas were later argued with addressing the differences~\cite{Pickett.prb2004,Millis.prl2011}. Very recently, superconductivity at $T_\mathrm{c}$ = 9-15 K was reported in the Nd$_{0.8}$Sr$_{0.2}$NiO$_2$ single-crystalline thin films deposited on a SrTiO$_3$ substrate~\cite{nature2019}. The finding makes the layered nickelates a hot research topic in the condensed matter community~\cite{Nd112.ZXS,Nd112.Norman,Pickett2020,Nd112.multiorbital,112.two-band.PRR2019,
112.band-engineering.PRB2019,R112.DFT.PRB2019-Jiang,prb2020.GMZhang,Nd112.YJL,t-j,whh}.

Most layered nickelates are structurally related to the Ruddlesden-Popper (RP) series, $L_{n+1}$Ni$_n$O$_{3n+1}$ ($L$ = lanthanide elements)~\cite{Honig.jssc1994}, which contains perovskite-type ($L$NiO$_3$)$_n$ block-layers that are connected with a rock-salt-type $L$O layer. The average formal Ni valence in $L_{n+1}$Ni$_n$O$_{3n+1}$ changes with $n$, being 2+, 2.5+, 2.67+, and 3+, respectively, at $n=1, 2, 3$, and $\infty$. Furthermore, the apical oxygen atoms in between the NiO$_2$ planes can be removed by a topochemical reduction, giving rise to the variant series $L_{n+1}$Ni$_n$O$_{2n+2}$~\cite{Lacorre.jssc1992} with the formal Ni valence of 1.5+, 1.33+, and 1+, respectively, for $n=2, 3$, and $\infty$. Note that the Ni valence state in the trilayer $L_4$Ni$_3$O$_{8}$ is mostly close to that in the superconducting nickelate Nd$_{0.8}$Sr$_{0.2}$NiO$_2$. Therefore, the trilayer nickelates seem to be promising to realize superconductivity in the bulk form.

As the precursor of $L_4$Ni$_3$O$_{8}$, $L_4$Ni$_3$O$_{10}$ belong to the RP nickelates which contains trilayers of NiO$_2$ sheets. So far, there are only three members in the family, with $L$ = La, Pr, and Nd, respectively. Zhang and Greenblatt~\cite{Greenblatt.jssc1995} earlier reported the synthesis, structure, and physical properties of $L_4$Ni$_3$O$_{10-\delta}$. Among them, La$_4$Ni$_3$O$_{10}$ showed a metallic behavior with Pauli paramagnetism. For $L$ = Pr and Nd, a metal-to-metal transition (MMT) was observed at 145 and 165 K, respectively, from the resistivity measurements. It was later indicated that the oxygen stoichiometry of La$_4$Ni$_3$O$_{10\pm\delta}$ influenced the electronic properties~\cite{Carvalho.43d.jap2000}. A similar MMT at 140 K was also observed for the as-prepared La$_4$Ni$_3$O$_{10.02}$ and reduced La$_4$Ni$_3$O$_{9.78}$, while the oxidized La$_4$Ni$_3$O$_{10.12}$ did not show evidence of an MMT. Recently, the MMT in La$_4$Ni$_3$O$_{10}$ was found to be accompanied with a structural response, featured with the expansion in the $b$ axis, but without any change in the space group~\cite{43d-SX.2019,43d.Kumar.jmmm2020}. Note that the temperature dependence of magnetic susceptibility only shows a gradual decrease at the MMT~\cite{PRB2001.La43d,Carvalho.43d.jap2000,43d.Kumar.jmmm2020}, apparently contradicting with the obvious jumps in the temperature dependence of resistivity and specific heat~\cite{43d-SX.2019,43d.Kumar.jmmm2020}.

The MMT was earlier attributed to a charge-density-wave (CDW) instability~\cite{Greenblatt.jssc1995} or charge ordering~\cite{Carvalho.43d.jap2000}. The tight-binding bandstructure calculation study suggested two hidden one-dimensional Fermi surfaces which could be responsible for the charge density wave instability~\cite{TBA.IC1996}. Recent angle resolved photoemission
spectroscopy (ARPES) measurements~\cite{43d.nc2017} on La$_4$Ni$_3$O$_{10}$ crystals indicated that, at the MMT, a gap of 20 meV opens in a flat band with strong $d_{3z^{2}-r^{2}}$ orbital character, whereas no pseudogap was found in the band with the dominant $d_{x^{2}-y^{2}}$ character. Nevertheless, the origin of the MMT remains elusive.

As the third member of $L_4$Ni$_3$O$_{10}$, Nd$_4$Ni$_3$O$_{10}$ has been rarely studied~\cite{Greenblatt.jssc1995,Nd43d-neutron.jssc2000}. Albeit of the resistivity jump, no magnetic anomaly was observable, primarily due to the large magnetic contributions from the Nd$^{3+}$ ions. Another motivation of this work is that Nd$^{3+}$ has the smallest ionic radius among the $L^{3+}$ ions~\cite{radius}, which gives rise to the smallest tolerance factor ($t$ = 0.932, 0.917 and 0.91, respectively, for La$_4$Ni$_3$O$_{10}$, Pr$_4$Ni$_3$O$_{10}$, and Nd$_4$Ni$_3$O$_{10}$~\cite{Greenblatt.jssc1995}). This would significantly influence the physical properties, like the case in the $L$NiO$_3$ system~\cite{RNiO3,Medarde1997}. In this paper we report the physical properties of Nd$_4$Ni$_3$O$_{10-\delta}$, particularly focusing on the MMT and the low-temperature properties. We found that the MMT is not only identified by the resistivity jump, but also characterized by a magnetic susceptibility drop, a specific-heat peak, and a nearly isotropic lattice contraction. Compared with the sister compound La$_4$Ni$_3$O$_{10}$, furthermore, the material shows a much larger electronic specific-heat coefficient and an enhanced log$T$ dependence in the low-temperature resistivity, indicating a novel heavy-fermion correlated electronic state in the title material.


\section{\label{sec:level2}Experimental methods}

Nd$_4$Ni$_3$O$_{10-\delta}$ polycrystalline samples were synthesized via high-temperature solid-state reactions. The source materials were Nd$_2$O$_3$ (99.997\%, Alfa Aesar) and NiO (99.998\%, Alfa Aesar), which were mixed in the stoichiometric ratio (Nd : Ni = 4 : 3). The ground mixture was first calcined at 1100 $^{\circ}$C in oxygen atmosphere, holding for 36 hours. The resultant was found to be Nd$_2$NiO$_{4}$, Nd$_3$Ni$_2$O$_{7-\delta}$, and Nd$_4$Ni$_3$O$_{10-\delta}$. In order to obtain the single-phase sample of Nd$_4$Ni$_3$O$_{10-\delta}$, the intermediate product (being reground and pressed into pellets) was sintered at 1100 $^{\circ}$C for 48 hours in oxygen atmosphere (0.3-0.5 MPa at 1100 $^{\circ}$C). The oxygen gas was generated by the decomposition of Ag$_2$O (99.7\%, Aladdin) and, together with the sample pellets, appropriate amount of Ag$_2$O was put and sealed in an evacuated silica ampule. This high-temperature solid-state reaction led to single-phase sample of Nd$_4$Ni$_3$O$_{10-\delta}$. According to the previous literatures~\cite{Greenblatt.jssc1995,Nd43d-neutron.jssc2000}, the sample prepared under an oxygen pressure of $\sim$ 0.1 MPa has oxygen deficiency with $\delta\sim$ 0.15. The as-prepared sample was found to be stable in air.

The sample was structurally examined by powder x-ray diffractions (XRD) using a PANalytical diffractometer (Empyrean Series 2) with a monochromatic Cu-$K_{\alpha1}$ radiation. The crystal structure were refined by a Rietveld analysis using the GSAS package~\cite{GSAS}. The fractional coordinates and the occupancies of the oxygen atoms were fixed according to the neutron diffraction result~\cite{Nd43d-neutron.jssc2000}, because neutron diffraction in general gives more accurate atomic positions for oxygen.

The temperature dependence of electrical resistivity and heat capacity was measured on a Quantum Design Physical Properties Measurement System (PPMS-9). In the resistivity measurement, the sample pellet was cut into a thin rectangular bar on which four parallel electrodes were made with silver paste. The dc magnetization was measured on a Quantum Design Magnetic Property Measurement System (MPMS3). The high-pressure resistivity was measured with the standard four-probe method in a palm-type cubic anvil cell (CAC) apparatus~\cite{hp.cjg}. Glycerol was used as the pressure transmitting medium. The pressure values were estimated from the pressure-load calibration curve determined by observing the characteristic phase transitions of Bi (2.55, 2.7, 7.7 GPa), Sn (9.4 GPa), and Pb (13.4 GPa) at room temperature. It should be noted that the pressure values inside the CAC exhibit slight variations upon cooling, which has been well characterized in the previous work~\cite{hp.cjg}.

\section{\label{sec:level3}Results and discussion}

\subsection{\label{subsec:level1}Crystal structure}
\begin{figure}
\includegraphics[width=8.5cm]{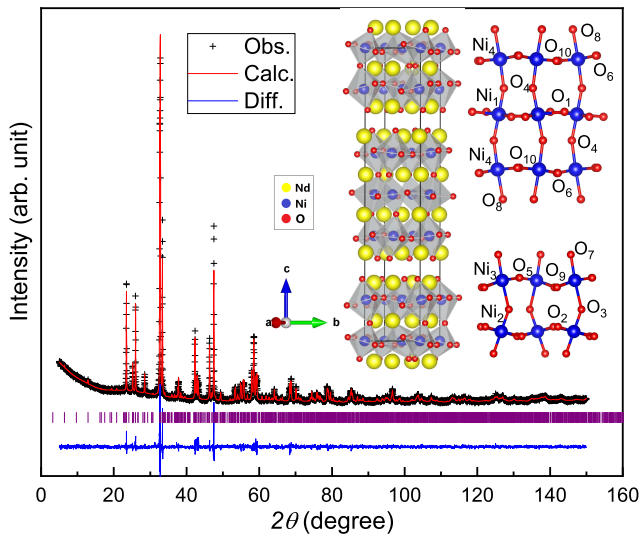}
\caption{Powder X-ray diffraction at room temperature and its Rietveld refinement profile of Nd$_4$Ni$_3$O$_{10-\delta}$. The insets show the crystal structure with vertex-sharing NiO$_6$ octahedra.}
\label{xrd}
\end{figure}

Figure~\ref{xrd} shows the XRD and its Rietveld-analysis profile for the Nd$_4$Ni$_3$O$_{10-\delta}$ sample studied in this paper. The Rietveld refinement was based on the structural model with the space group of $P2_1/a$ and $Z$ = 4~\cite{Nd43d-neutron.jssc2000}. The $R$ factors and the goodness of the refinement are $R_{\mathrm{wp}}$ = 5.0\%, $R_{\mathrm{p}}$ = 3.47\%, and $\chi^{2}$ = 1.94, respectively, suggesting reliability of the structural parameters fitted. The unit-cell parameters obtained are $a=5.36550(6)$ {\AA}, $b=5.45462(6)$ {\AA}, $c=27.4186(3)$ {\AA}, and $\beta=90.318(1)^{\circ}$, which are consistent with the previous report ($a=5.3675(2)$ {\AA}, $b=5.4548(2)$ {\AA}, $c=27.433(1)$ {\AA}, and $\beta=90.312(2)^{\circ}$~\cite{Nd43d-neutron.jssc2000}).

Shown in the inset of Fig.~\ref{xrd} is the crystal structure of Nd$_4$Ni$_3$O$_{10-\delta}$, which contains triple perovskite-type block layers in which the vertex-sharing NiO$_6$ octahedra are distorted, twisted and tilted. There are four distinct crystallographic sites for Ni atoms, two of them are in the inner layer (IL), and others in the outer layer (OL). To assess the possible charge ordering state of Ni$^{3+}-$Ni$^{2+}-$Ni$^{3+}$ in the trilayer~\cite{Carvalho.43d.jap2000}, we calculated the bond valence sum (BVS)~\cite{bvs} for the Ni ions, using the formula $\sum$ exp$(\frac{R_{0}-d_{ij}}{0.37})$, where $R_{0}$ is 1.654 \r{A} for a Ni$^{2+}-$O bond and $d_{ij}$ are the measured interatomic distances between Ni cations and the coordinated oxygen anions. As is seen in Table~\ref{bvs}, the BVS values of the Ni atoms are about 2.6, consistent with the average formal Ni valence in Nd$_4$Ni$_3$O$_{10}$. However, there is no tendency of the charge ordering of Ni$^{3+}-$Ni$^{2+}-$Ni$^{3+}$. On the contrary, the Ni valence in the IL turns out to be higher. Similar results were reported for La$_4$Ni$_3$O$_{10}$ and Pr$_4$Ni$_3$O$_{10}$~\cite{43d-SX.2019}. Therefore, the possible extreme charge ordering scenario should be the case that the Ni valence in the IL is Ni$^{3+}$ and, correspondingly, the formal Ni valence in the OLs could be 2.5+ for the stoichiometric Nd$_4$Ni$_3$O$_{10}$. Note that the interatomic distances between Ni and the apical oxygen, $d_{\mathrm{ap}}$, are very different in the ILs and OLs. In the OLs, $d_{\mathrm{ap}}$ are obvious larger, reflecting a Jahn-Teller-like distortion or an orbital polarization~\cite{OP}. The higher Ni valence in the ILs seems to be related to the shorter $d_{\mathrm{ap}}$. It is of great interest to have a similar analysis for the low-temperature crystallographic data using neutron diffractions (such a work is under way). By the way, we also calculate the BVS values for the Nd ions. As listed in Table~\ref{bvs}, they are reasonably close to the conventional valence of Nd$^{3+}$, albeit of the difference in the coordination number.

\begin{table*}
\caption{Bond valence sums (BVS) of Nd and Ni calculated with the related interatomic distances at 300 K in Nd$_4$Ni$_3$O$_{10-\delta}$. IL and OL are the abbreviations for inner layer and outer layer, respectively.}
  \label{bvs}\renewcommand\arraystretch{1.3}
  \begin{tabular}{cccccccccccccccccc}
      \hline \hline
  Nickel  & site   & $x$ & $y$ & $z$ & &\multicolumn{2}{c}{$d$(Ni$-$O$ _{\mathrm{ap}}$)}&& \multicolumn{4}{c}{$d$(Ni$-$O$ _{\mathrm{eq}}$)}&&\multicolumn{3}{c}{BVS(Ni)} \\
  \hline
    Ni1 (IL1)      & 2b  & 0.0000 & 0.5000 & 0.5000 & &1.945 & 1.945 && 1.946& 1.979  & 1.946 &1.979 &&\multicolumn{3}{c}{2.64} \\
    Ni2 (IL2)      & 2a &0.0000 & 0.0000 & 0.0000&& 1.936  & 1.936 &&1.949 & 1.949  & 1.932 &1.932&&\multicolumn{3}{c}{2.77}  \\
    Ni3 (OL1)      & 4e  &-0.0157 & 0.0146 & 0.1406 && 1.994  & 2.117 &&1.956 & 2.042  & 1.867 &1.877&&\multicolumn{3}{c}{2.58}  \\
    Ni4 (OL2)     & 4e &0.4988 & 0.0123 & 0.6409&& 2.118 & 1.989 && 1.882& 1.897  & 1.993 &1.951&&\multicolumn{3}{c}{2.59} \\
    \hline
    Neodymium  & site  & $x$ & $y$ & $z$ &\multicolumn{4}{c}{$d$(Nd$-$O$ _{\mathrm{up}}$)} &\multicolumn{4}{c}{$d$(Nd$-$O$ _{\mathrm{mid}}$)}&\multicolumn{4}{c}{$d$(Nd$-$O$ _{\mathrm{dn}}$)}& BVS(Nd) \\
    \hline
    Nd1 (OL1)      & 4e  &-0.0062 & 0.0177 & 0.3010 &\multicolumn{4}{c}{2.343}  &2.832&2.350&2.639&3.194&2.325 &2.652&2.472&2.417& 2.89 \\
    Nd2 (OL2)     & 4e  &0.5002 & 0.0111 & 0.8004 & \multicolumn{4}{c}{2.335}  &3.106&2.897&2.440&2.567&2.399 &2.391&2.700&2.701& 2.58 \\
    Nd3 (IL1)     & 4e  &0.0369 & 0.0069 & 0.4312 &2.626&3.123&2.862&2.588 &2.961&2.373&2.476&3.103&2.345 &3.250&2.560&2.681& 2.71 \\
    Nd4 (IL2)      & 4e  &0.5153 & -0.0092 & 0.9309 & 2.506 &2.859&2.573&2.919 &2.263&2.772&3.192&2.642&2.759 &2.578&3.025&2.462& 2.73 \\
    \hline
    \hline
  \end{tabular}
\end{table*}

\subsection{\label{subsec:level2}Electrical resistivity}
\begin{figure}
\includegraphics[width=8.5cm]{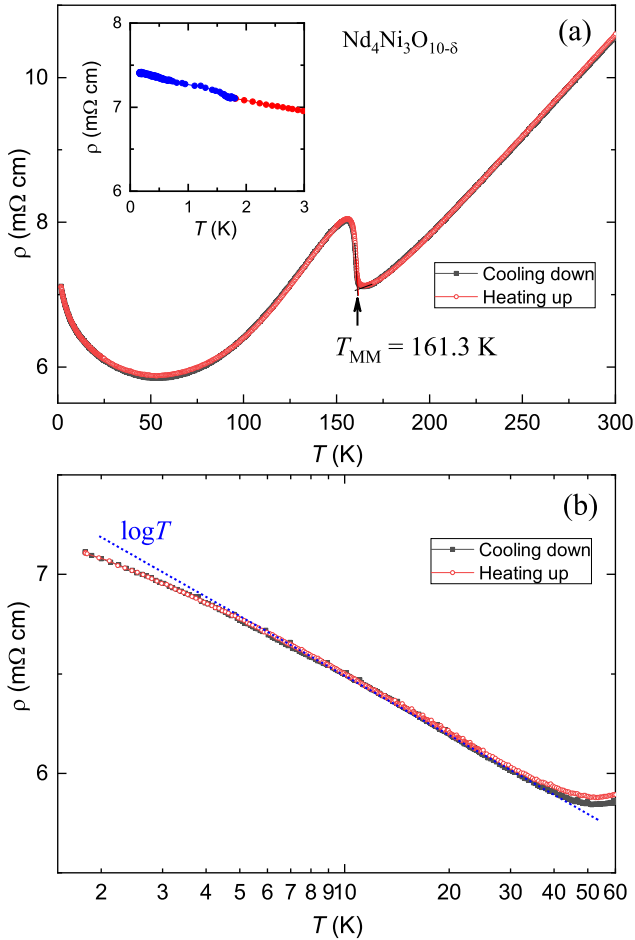}
\caption{Temperature (a) and logarithmic temperature (b) dependence of electrical resistivity of the Nd$_4$Ni$_3$O$_{10-\delta}$ polycrystalline sample.}
\label{rt}
\end{figure}

Figure~\ref{rt} shows the temperature dependence of resistivity for the as-prepared Nd$_4$Ni$_3$O$_{10-\delta}$ polycrystalline sample. The $\rho(T)$ behavior is basically metallic, and no sign of superconductivity is observed down to 0.16 K [see the inset of Fig.~\ref{rt}(a)]. One can immediately see a resistivity jump at $T_{\mathrm{MM}}$ = 161.3 K, indicating a MMT which is basically consistent with the previous observation (the transition temperature, defined as the onset of the steep resistivity increase, was actually $\sim$155 K rather than 165 K in the previous report~\cite{Greenblatt.jssc1995}). No thermal hysteresis is obvious, suggesting a second-order transition or a weakly first-order transition. The finite resistivity jump implies a partial bandgap opening at the Fermi level, $E_{\mathrm{F}}$.

Noticeably, the $\rho(T)$ data show an obvious upturn below 50 K, which approximately obeys a logarithmic temperature dependence, as shown in Fig.~\ref{rt}(b). The result is different from that of its sister compound La$_4$Ni$_3$O$_{10-\delta}$, the latter of which only shows a tiny (if not none) resistivity upturn ~\cite{Greenblatt.jssc1995,PRB2001.La43d,Carvalho.43d.jap2000,43d.nc2017,43d.Kumar.jmmm2020}. In comparison, the low-temperature $\rho(T)$ curve of Pr$_4$Ni$_3$O$_{10.1}$ exhibits a clearer upturn~\cite{Pr43d}. This trend suggests that the enhanced resistivity upturn in Nd$_4$Ni$_3$O$_{10-\delta}$ is associated with either the magnetism of Nd$^{3+}$ ions or the smaller Nd$^{3+}$ ions (compared with La$^{3+}$). The latter gives rise to a strong lattice distortion and a consequent ``more localized" electronic state akin to the case in $L$NiO$_3$~\cite{RNiO3,Medarde1997}. In general, the energy level of Nd-4$f$ electrons is far below the $E_{\mathrm{F}}$, and the effective hybridization with the conduction bands is negligible. Note that a similar logarithmic temperature dependence of resistivity was seen in NdNiO$_{2}$~\cite{nature2019} and LaNiO$_{2}$~\cite{La112.Ikeda2016} thin films, which was recently interpreted in terms of the Ni-moment centered Kondo scattering~\cite{prb2020.GMZhang}. Such a novel Kondo-like interaction was also theoretically discussed~\cite{Pickett2020,t-j} and was recently demonstrated by the x-ray spectroscopy and density functional calculations for in NdNiO$_{2}$ or LaNiO$_{2}$~\cite{Nd112.ZXS}. For Nd$_4$Ni$_3$O$_{10-\delta}$ here, partial Ni-3$d$ electrons possibly become localized below $T_{\mathrm{MM}}$ (akin to the site-selective Mottness~\cite{Nd113.Mottness,SSMott.prl2012}), which carry magnetic moments (see the following analysis). Such magnetic moments could serves as the Kondo-scattering centers, like the case in NdNiO$_{2}$ or LaNiO$_{2}$, which may give rise to the low-temperature resistivity upturn.

\subsection{\label{subsec:level3}Magnetic properties}

\begin{figure}
\includegraphics[width=8.5cm]{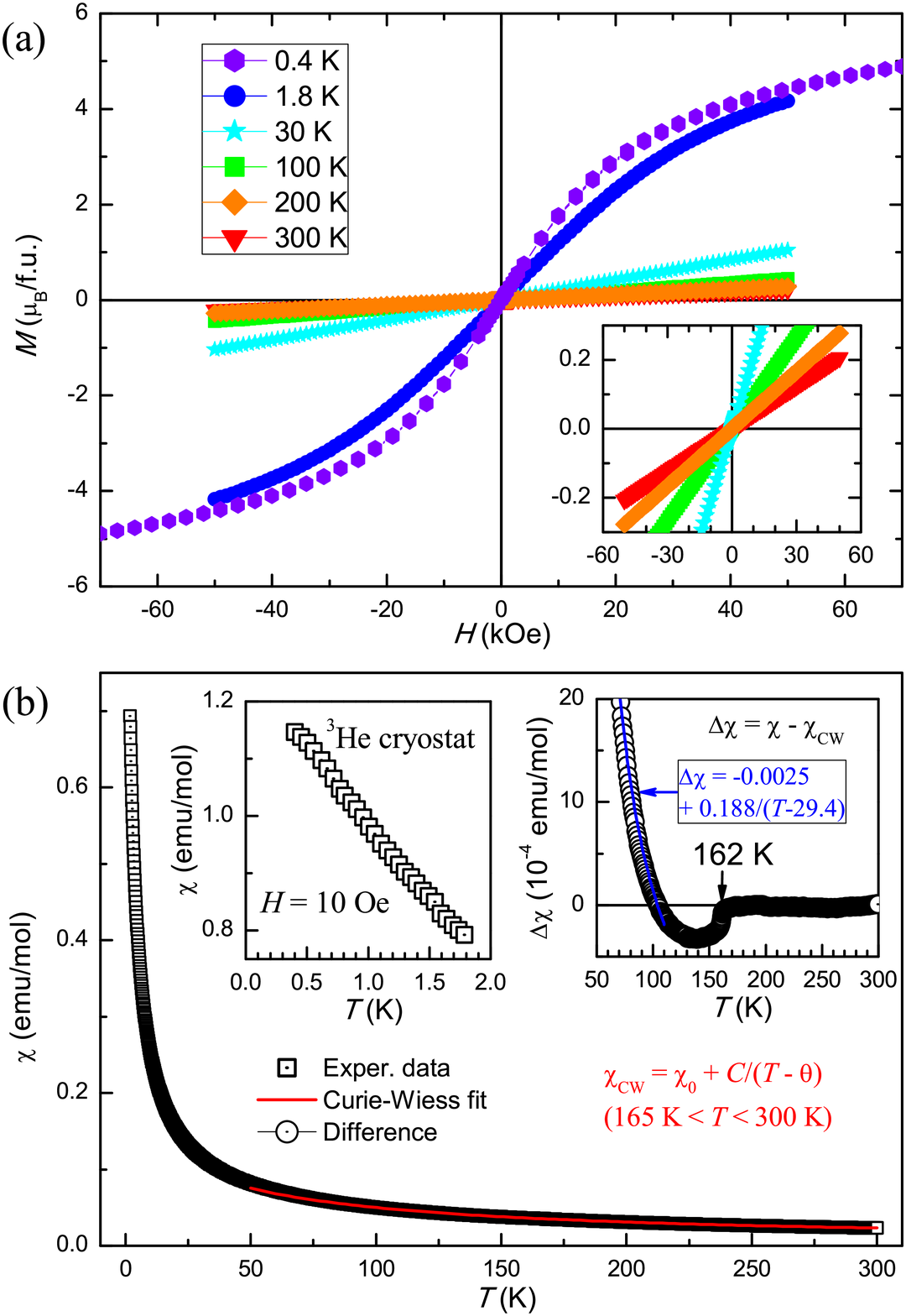}
\caption{(a) Magnetic field dependence of magnetization at some fixed temperature for Nd$_4$Ni$_3$O$_{10-\delta}$. The inset is a close-up for the high-temperature data, showing the linear dependence. (b) Temperature dependence of magnetic susceptibility under a magnetic field of $H$ = 10 kOe for Nd$_4$Ni$_3$O$_{10-\delta}$. The left inset shows the susceptibility down to 0.45 K using a He-3 cryostat. The right inset plots the result of susceptibility subtraction with a Curie-Weiss fit.}
\label{Mag}
\end{figure}

The magnetic measurement data of Nd$_4$Ni$_3$O$_{10-\delta}$ were shown in Fig.~\ref{Mag}. First of all, the field dependence of magnetization is essentially linear at $T>$ 30 K, indicating that the sample is free of ferromagnetic impurities. At 1.8 K, the $M(H)$ curve is Brillouin function like and, at the lowest temperature down to 0.4 K, the magnetization almost saturates at $\sim$ 5 $\mu_\mathrm{B}$ per formula unit (fu). The magnetic moment is severely reduced for the Nd$^{3+}$ ions, since the theoretical ordered moment of a free Nd$^{3+}$ ion is $g_{J}J$ = 3.27 $\mu_\mathrm{B}$, equivalent to 13 $\mu_\mathrm{B}$/fu for Nd$_4$Ni$_3$O$_{10-\delta}$. The reduction of the Nd$^{3+}$ moment is commonly attributed to the crystalline-electric-field (CEF) effect. The CEF effect often gives rise to a $J_{\mathrm{eff}}$ = 1/2 ground state for an odd number of 4$f$ electrons. Indeed, the following specific-heat measurement confirms this scenario.

As shown in Fig.~\ref{Mag}(b), the temperature dependence of susceptibility of Nd$_4$Ni$_3$O$_{10-\delta}$ is Curie-Weiss (CW) like. The CW fit in the temperature range of 165 K $<T<$ 300 K using the formula $\chi=\chi_0+C/(T+\theta_\mathrm{W})$ yields a temperature-independent term $\chi_0$ = 0.0044 emu/mol-fu, a Curie constant $C$ = 6.43 emu K/mol-fu, and a paramagnetic Weiss temperature $\theta_\mathrm{W}$ = 40.5 K. With the fitted Curie constant, the effective local magnetic moment is derived to be 3.59 $\mu_\mathrm{B}$/Nd$^{3+}$, close to the theoretical value of Nd$^{3+}$ free ions (3.62 $\mu_\mathrm{B}$). This suggests that the magnetic moment from Ni at $T>T_{\mathrm{MM}}$, if exists, is negligible. In spite of a significantly high value of $\theta_\mathrm{W}$, which means substantial antiferromagnetic interactions between the Nd$^{3+}$ moments, no magnetic transition associated with the Nd$^{3+}$ moment is observed above 0.4 K. This could be due to a frustration effect, since the dominant magnetic coupling seems to be an indirect RKKY interaction. Finally, $\chi_0$ is remarkably larger than the $\chi$ value at 300 K for La$_4$Ni$_3$O$_{10}$ (0.0018 emu/mol-fu)~\cite{Greenblatt.jssc1995,Carvalho.43d.jap2000}. The large value $\chi_0$ should be mostly contributed from the exchange-enhanced Pauli paramagnetism and Van Vleck paramagnetism.

To detect a possible change in the magnetic susceptibility at the MMT, we made a subtraction using the CW-fit data as the reference. As shown in the right inset of Fig.~\ref{Mag}(b), the subtraction reveals a susceptibility drop of $\sim3\times 10^{-4}$ emu/mol-fu. The susceptibility drop corresponds to a $N(E_\mathrm{F})$ loss of $\sim$ 9 states/eV, suggesting a partial gap opening. Note that the susceptibility drop is so far exclusively observed for the MMT in the $L_4$Ni$_3$O$_{10}$ family (only a gradual decrease at the MMT was seen for La$_4$Ni$_3$O$_{10}$~\cite{PRB2001.La43d,Carvalho.43d.jap2000,43d.Kumar.jmmm2020}). Also noted is that, after the subtraction, the $\chi(T)$ data below 140 K still show a CW-like behavior. This remaining CW term, with an effective local moment of $\sim$ 1.2 $\mu_\mathrm{B}$/fu (fitted with the data from 50 K to 120 K), could arise from the partial localized electronic states of Ni-3$d$ electrons. This unusual state resembles the site-selective Mottness in $L$NiO$_3$~\cite{SSMott.prl2012}. As mentioned above, the Ni in the IL shows a higher BVS value, which implies dominant Ni$^{3+}$ oxidation states in the ILs. Remember that NdNiO$_3$ with Ni$^{3+}$ oxidation states shows a metal-to-insulator transition at 201 K~\cite{RNiO3}. Thus the Ni-3$d$ electronic states in the ILs are likely to be localized below $T_{\mathrm{MM}}$. If this is the case, the MMT in the present system involves an interlayer charge ordering, in addition to the possible charge-transfer gap earlier proposed for $L$NiO$_3$~\cite{RNiO3,Medarde1997}.

\subsection{\label{subsec:level4}Specific heat}

\begin{figure*}
\includegraphics[width=15cm]{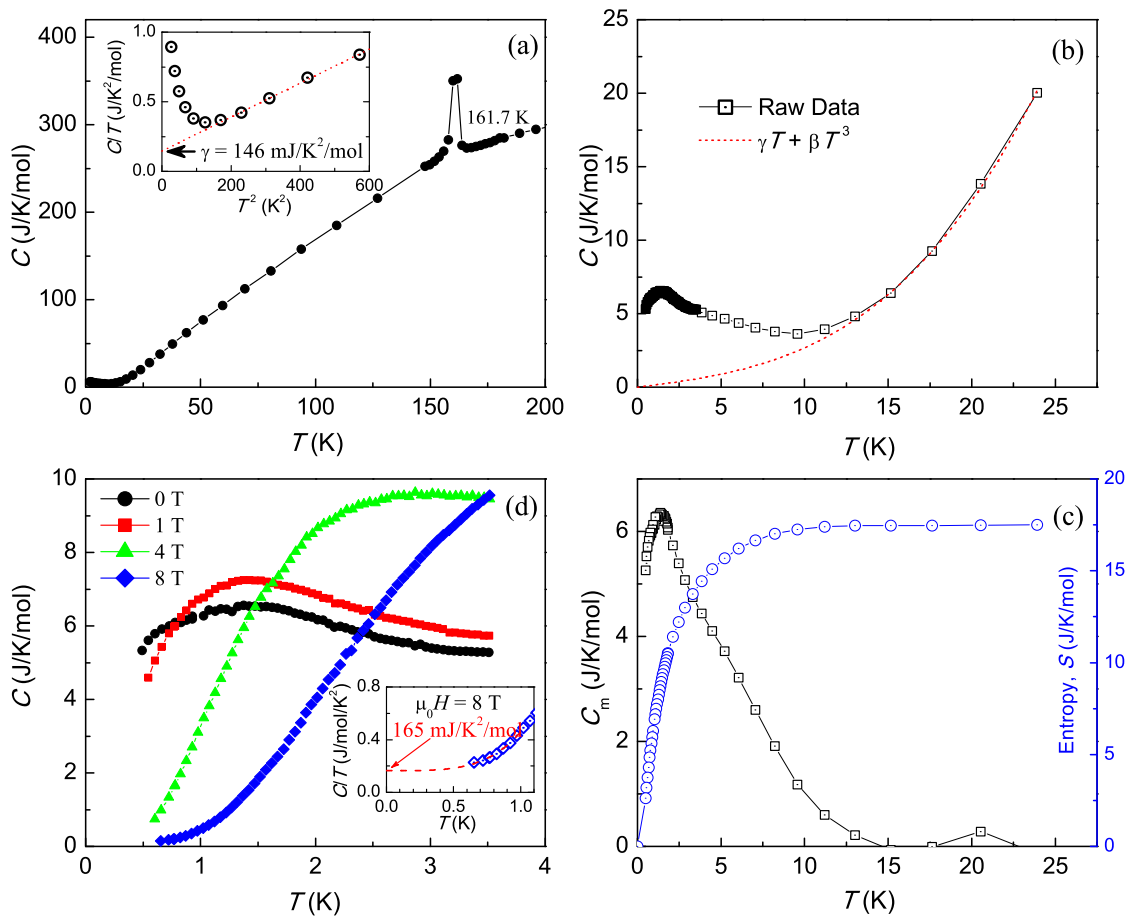}
\caption{Temperature dependence of specific heat for Nd$_4$Ni$_3$O$_{10-\delta}$. The insets of (a) and (d) plot $C/T$ as functions of $T^2$ and $T$, respectively, for extracting the electronic specific-heat coefficient. (c) Magnetic contributions of specific heat separated by the subtraction shown in (b). The right axis of (c) shows the magnetic entropy.  }
\label{sh}
\end{figure*}

Figure~\ref{sh} shows the temperature dependence of specific heat of the Nd$_4$Ni$_3$O$_{10-\delta}$ sample. Remarkably, there is a sharp peak at 161 K, further confirming the bulk nature of the MMT. The entropy associated with the MMT is extracted to be 2.8 J K$^{-1}$ mol-fu$^{-1}$, which is about twice of that in La$_4$Ni$_3$O$_{10}$~\cite{43d.Kumar.jmmm2020}. At $T<$ 10 K, the specific heat shows an upturn, as is clearly seen in the plot of $C/T$ vs. $T^2$ [inset of Fig.~\ref{sh}(a)]. In the temperature range of 13 K $<T<$ 25 K, a linear relation between $C/T$ and $T^2$ exists, reflecting the dominant contributions from the electronic part ($\gamma T$) and the phonon part ($\beta T^3$). The linear fit yields $\gamma$ = 146 mJ K$^{-2}$ mol-fu$^{-1}$ and $\beta$ = 1.22 mJ K$^{-4}$ mol-fu$^{-1}$. The $\gamma$ value is almost one order of magnitude lager than that in La$_4$Ni$_3$O$_{10}$~\cite{PRB2001.La43d,43d.Kumar.jmmm2020}. This would suggest that the thermal effective mass $m^*/m_0\sim$ 26~\cite{PRB2001.La43d,43d.Kumar.jmmm2020}, suggesting a heavy-electron behavior. Finally, with the formula $\theta_{\mathrm{D}}=[(12/5)NR\pi^4/\beta]^{1/3}$, where $N$ denotes the number of atoms per formula unit ($N$ = 17), the Debye temperature $\theta_{\mathrm{D}}$ was estimated to be 300 K. This $\theta_{\mathrm{D}}$ value is in between those of the previous reports (256 K~\cite{PRB2001.La43d} and 384 K~\cite{43d.Kumar.jmmm2020}) for La$_4$Ni$_3$O$_{10}$.

To clarify the specific-heat tail below 10 K, we carried out the specific-heat measurement down to 0.5 K using a He-3 cryostat. The data are shown in Figs.~\ref{sh}(b,d). There is a broad hump at 0.5-2 K, which is commonly attributed to the Schottky anomaly from the Nd$^{3+}$ ions~\cite{NdMnO3,Schottky.LXG}. We were able to separate the magnetic contributions by removing the electronic and phonon parts, which are plotted in Fig.~\ref{sh}(c). One sees an additional shoulder at around 5 K. This could be due to a complex Schottky anomaly (because of the complex CEF) and/or a possible Kondo effect~\cite{kondo.HC}. The total magnetic entropy is 17.5 J K$^{-1}$ mol$^{-1}$, not far from an expected value of 4$R$ln2 = 23.1 J K$^{-1}$ mol$^{-1}$ for a $J_{\mathrm{eff}}=$1/2 ground state. The result is basically consistent with the low-lying doublet splitting of Nd$^{3+}$ ions.

Under magnetic fields, the broad bump shifts to higher temperatures, as shown in Fig.~\ref{sh}(d), further confirming the Schottky anomaly scenario~\cite{NdMnO3,Schottky.LXG}. The specific heat at the lowest temperature of $\sim$ 0.5 K is mostly suppressed under a magnetic field of 8 T. This allows us to extract the electronic specific-heat coefficient independently. We thus made a polynomial extrapolation (roughly assuming that the magnetic contribution at 8 T is  $\delta T^5$) down to zero temperature, which yields $(C/T)_{T\rightarrow 0}\approx$ 165 mJ K$^{-2}$ mol-fu$^{-1}$ [see the inset of Fig.~\ref{sh}(b)]. The result well agrees with the $\gamma$ value obtained above.

\subsection{\label{subsec:level5}Low-temperature XRD}
\begin{figure*}
\includegraphics[width=15cm]{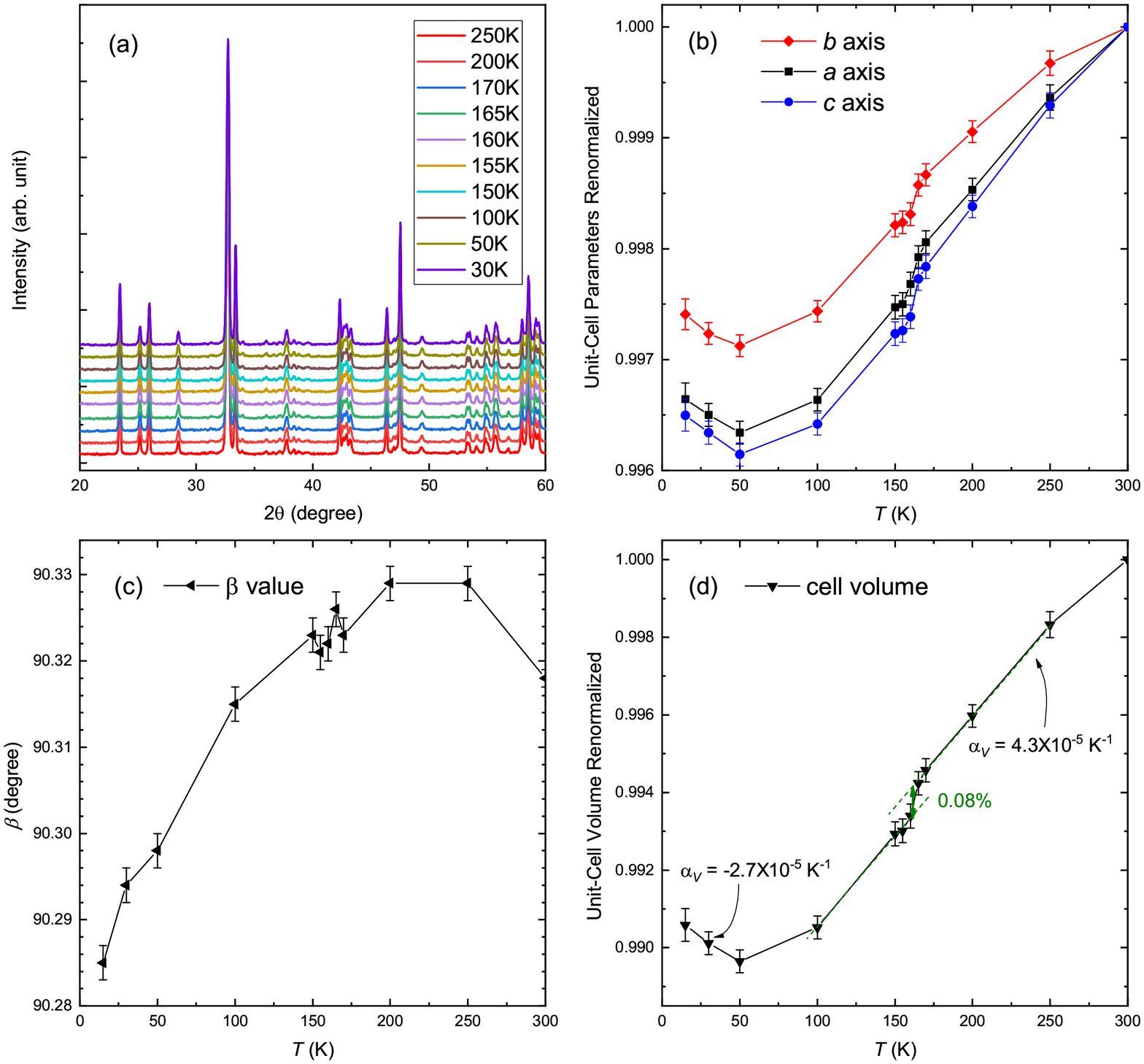}
\caption{(a) Low-temperature X-ray diffractions for Nd$_4$Ni$_3$O$_{10-\delta}$. (b) Temperature dependence of renormalized lattice parameters $a$, $b$, and $c$, which shows a lattice contraction at 162 K and lattice expansions below 50 K. (c,d) Temperature dependence of the $\beta$ value and the renormalized unit-cell volume, respectively.}
\label{ltxrd}
\end{figure*}

To examine the possible structure change at the MMT, we conducted the low-temperature XRD measurement on Nd$_4$Ni$_3$O$_{10-\delta}$. The XRD data at 15 K can be well fitted on the basis of the identical crystal structure with the space group of $P2_1/a$. And we cannot distinguish the monoclinic-II phase ($Z$ = 2) from the the monoclinic-I phase ($Z$ = 4)~\cite{43d.structure.PRB2018}. Fig.~\ref{ltxrd}(a) shows the XRD patterns at various temperatures. They are essentially the same except for slight peak shifts due to the changes in lattice parameters. The subtraction of the XRD data between 150 K and 170 K does not show any additional Bragg peaks. A similar result was reported in Rb$_{1-\delta}$V$_2$Te$_2$O~\cite{1221V.Ablimit} and La$_4$Ni$_3$O$_{10}$ using synchrotron radiations~\cite{43d.Kumar.jmmm2020,43d-SX.2019}. Nevertheless, the possible CDW ordering cannot be ruled out due to the limitations of the XRD technique, and electron diffractions may be useful to clarify this issue in the future.

The lattice parameters were obtained by the Rietveld analysis, and its temperature dependence is shown in Figs.~\ref{ltxrd}(b-d). At $T\geq$ 170 K, the lattice parameters decrease almost linearly with decreasing temperature. The volume expansion coefficient, $\alpha_V$ = (1/$V$)($\partial V$/$\partial T$)$_P$, is $4.3\times$10$^{-5}$ K$^{-1}$. At $T=$ 162 K, a steep decrease is seen in all the unit-cell dimensions, resulting in a cell volume contraction of 0.08\% at the MMT. The result is different with that of La$_4$Ni$_3$O$_{10}$ in which only the $b$ axis shows a 0.02\% increase at the MMT~\cite{43d.Kumar.jmmm2020,43d-SX.2019}. The different structural response seems to be originated with the smaller tolerance factor in Nd$_4$Ni$_3$O$_{10-\delta}$.

Note that the lattice parameters \emph{increase} with decreasing temperature below 50 K, indicating an anomalous negative thermal expansion (NTE) with an $\alpha_V$ value of $-2.7\times$10$^{-5}$ K$^{-1}$. There are different types of mechanisms that may lead to a NTE~\cite{NTE.JCPM2005,NTE.CSR2015}. In the present case, we note that the NTE phenomenon happens coincidently with the resistivity upturn and even with the specific-heat upturn as described above. Thus the NTE is possibly associated with the ``heavy-Fermion" behavior as well as the Schottky contributions~\cite{NTE.JCPM2005}.

\subsection{\label{subsec:level6}High-pressure study}

\begin{figure*}
\includegraphics[width=14cm]{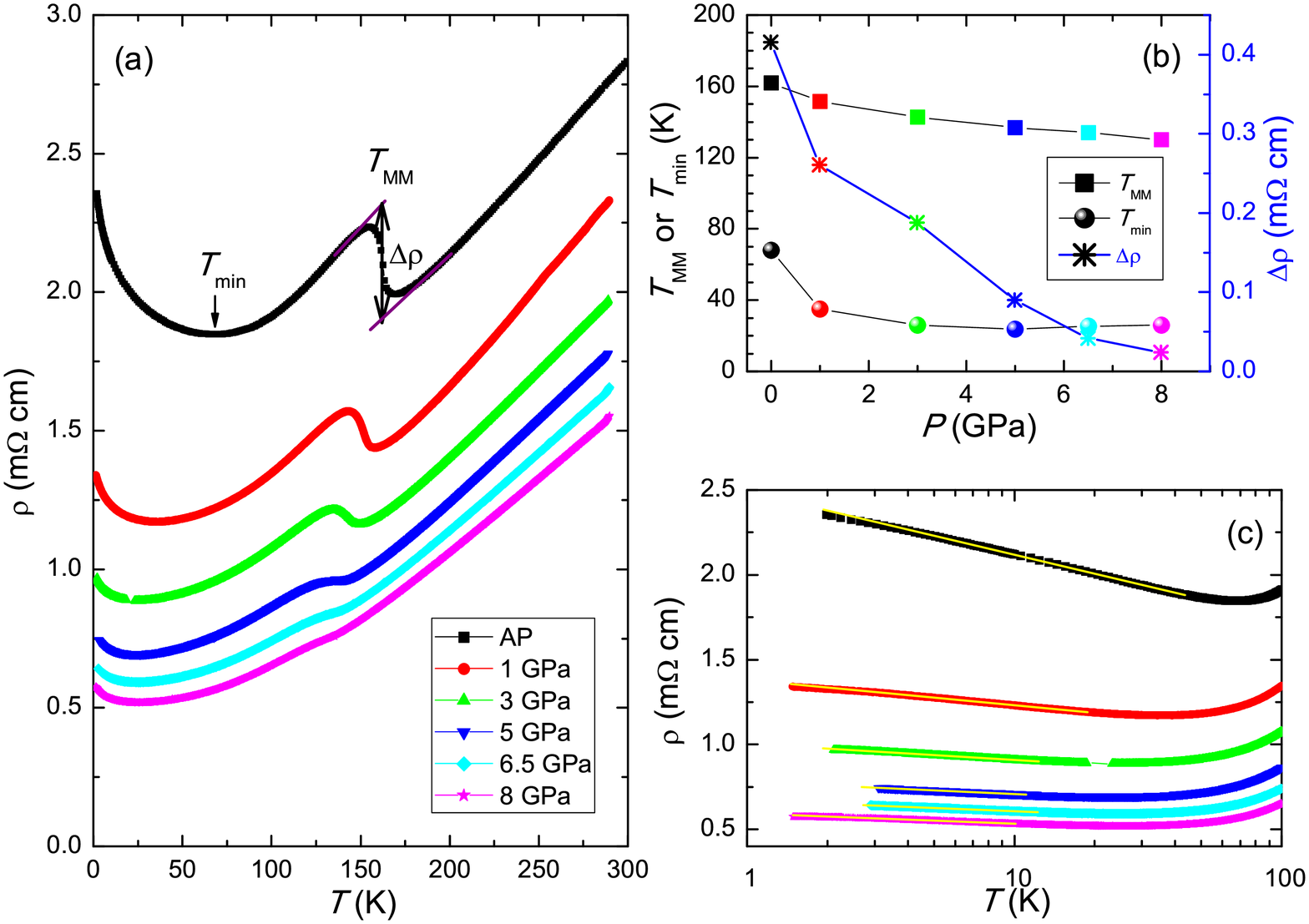}
\caption{(a) Temperature-dependent resistivity of Nd$_4$Ni$_3$O$_{10-\delta}$ under various hydrostatic pressures up to 8 GPa. (b) The metal-to-metal transition temperature ($T_{\mathrm{MM}}$), the resistivity jump at $T_{\mathrm{MM}}$ ($\Delta \rho$, right axis), and the temperature at which the resistivity shows a minimum ($T_{\mathrm{min}}$) are plotted as functions of pressure. (c) Logarithmic temperature dependence of resistivity at low temperatures and under pressures. The yellow straight lines are guides to the eyes.}
\label{hp}
\end{figure*}

To study the pressure effect on the MMT and the low-temperature Kondo-like behavior in Nd$_4$Ni$_3$O$_{10-\delta}$, we have measured the temperature-dependent resistivity under various hydrostatic pressures up to 8 GPa using another single-phase sample. As shown in Fig.~\ref{hp}(a), the $\rho(T)$ of Nd$_4$Ni$_3$O$_{10-\delta}$ at ambient pressure displays a jump at $T_{\mathrm{MM}}\approx$ 162 K. Above the $T_{\mathrm{MM}}$, the resistivity increases linearly with temperature, while below $T_{\mathrm{MM}}$ a broad resistivity minimum centered at $T_{\mathrm{min}}\sim$ 70 K appears with the resistivity upturn following a log$T$ dependence at low temperatures. Fig.~\ref{hp}(b) plots the $T_{\mathrm{MM}}$, the resistivity jump $\Delta \rho$, and the $T_{\mathrm{min}}$ as functions of pressure $P$. With increasing $P$, $T_{\mathrm{MM}}$ decrease mildly, from 162 K at 0 GPa to 130 K at 8 GPa with an initial slope of d$T$/d$P\approx -10$ K/GPa, comparable to that of La$_4$Ni$_3$O$_{10}$ ($-6.9$ K/GPa)~\cite{PRB2001.La43d}. The gradual suppression of $T_{\mathrm{MM}}$ with pressure in Nd$_4$Ni$_3$O$_{10-\delta}$ is also consistent with the observed lattice contraction at $T_{\mathrm{MM}}$ shown in Fig.~\ref{ltxrd}, since high pressure usually stabilizes the smaller-volume phase. Meanwhile $T_{\mathrm{min}}$ deceases rapidly at low pressures. And it tends to remain unchanged ($\sim$26 K) at higher pressures similar to the low-temperature $\rho(T)$ behavior of La$_4$Ni$_3$O$_{10}$ ($T_{\mathrm{min}}=$ 20 K)~\cite{43d.Kumar.jmmm2020}. The resistivity jump $\Delta \rho$ is mostly sensitive to pressure, and it diminishes much faster than $T_{\mathrm{MM}}$ and $T_{\mathrm{min}}$ do. From these trends, it is very likely that the MMT will be eliminated completely at a finite temperature without approaching a quantum critical point.

Fig.~\ref{hp}(c) shows a semi-logarithmic plot for the high-pressure low-temperature resistivity data. At ambient pressure the $\rho(T)$ data exhibit a logarithmic-temperature dependence below 50 K. At $P\geq$ 1 GPa, the Kondo-like resistivity upturn is remarkably reduced, coincident with the suppression of the MMT. The result suggests that the MMT may be the prerequisite of the Kondo-like electron-correlated behavior, in line with the Ni$^{3+}$ centered Kondo effect mentioned above. Note that the weak upturn of $\rho(T)$ at higher pressures is very similar to that of La$_4$Ni$_3$O$_{10}$~\cite{43d.Kumar.jmmm2020}. This suggests that the heavy-electron behavior in Nd$_4$Ni$_3$O$_{10-\delta}$ disappears largely under high pressures.

\subsection{\label{subsec:level7}Discussion}

From the results above, one sees that the MMT in Nd$_4$Ni$_3$O$_{10-\delta}$ bears both similarities and differences with that of its sister compound La$_4$Ni$_3$O$_{10}$. Both materials show a resistivity jump, a specific-heat anomaly, and a structural response at the MMT. Nevertheless, unlike La$_4$Ni$_3$O$_{10}$ which shows a continuous expansion in the $b$ axis at the MMT~\cite{43d-SX.2019,43d.Kumar.jmmm2020}, Nd$_4$Ni$_3$O$_{10-\delta}$ exhibits a nearly isotropic lattice contraction. Also Nd$_4$Ni$_3$O$_{10-\delta}$ exclusively exhibits an obvious magnetic susceptibility drop, indicating a significant loss of $N(E_\mathrm{F})$. The result suggests different types of charge ordering and/or CDW among the trilayer nickelate family, primarily due to the different tolerance factor as mentioned above. The similar ``ionic size effect" exists in perovskite-type nickelates $L$NiO$_3$~\cite{RNiO3}. An additional work related is that, in ultrathin LaNiO$_3$ films, the strain from the substrates governs the lattice distortion, which produces an emergent charge-ordered ground state~\cite{La113-films}.

Owing to the higher Ni valence in the IL of the (NdNiO$_3$)$_3$ trilayers, as indicated from the BVS value, the inner perovskite-like layer may undergo a metal-to-insulator transition like the case in NdNiO$_3$~\cite{RNiO3,Nd113.Mottness,SSMott.prl2012}. Meanwhile, the OLs possibly remain to be metallic. The resistivity jump and the magnetic susceptibility drop at the MMT support this picture. Additionally, a very recent work on Pr$_4$Ni$_3$O$_{10}$~\cite{Pr43d.2020} revealed metallic and semiconducting behaviors just below the MMT, respectively, for electric current flowing in the $ab$ plane and along the $c$ axis. Considered the similarity between Pr$_4$Ni$_3$O$_{10}$ and Nd$_4$Ni$_3$O$_{10}$, a similar anisotropy probably exists in Nd$_4$Ni$_3$O$_{10}$ also, which corroborates that the electronic state in the IL becomes largely localized below the MMT. In this context, the low-temperature state of Nd$_4$Ni$_3$O$_{10}$ may be a natural metal/insulator/metal superlattice of nickelate, first proposed by Chaloupka and Khaliullin~\cite{Khaliullin.prl2008}, in which orbital order and possible superconductivity were expected.

Importantly, the dominant Ni$^{3+}$ ions in the insulating ILs carry a magnetic moment, which may serve as the Kondo-scattering centers. Note that the itinerant conduction electrons are also from the Ni-3$d$ electrons (yet in the OLs). The situation is something like the site-selective Mott phase in $L$NiO$_3$ in which the $d$ electrons on part of the Ni$^{3+}$ ions are localized while the $d$ electrons on other Ni$^{3+}$ ions form a singlet with holes on the surrounding oxygen ions.~\cite{SSMott.prl2012}. With this picture in mind, the logarithmic temperature dependence of resistivity below 50 K and the heavy-electron behavior in Nd$_4$Ni$_3$O$_{10-\delta}$ can be understood in terms of the novel Kondo scattering.

\section{\label{sec:level4}Concluding remarks}

In summary, the trilayer nickelate Nd$_4$Ni$_3$O$_{10-\delta}$ exhibits a metal-to-metal transition at 162 K, featured by a resistivity jump, a susceptibility drop, a specific-heat peak, and a nearly isotropic lattice contraction. The magnetic susceptibility drop is observed for the first time in the trilayer nickelate family. Besides, Nd$_4$Ni$_3$O$_{10-\delta}$ shows a distinguishable logarithmic temperature dependence below 50 K. Furthermore, compared with La$_4$Ni$_3$O$_{10}$, Nd$_4$Ni$_3$O$_{10-\delta}$ possesses a much larger electronic specific-heat coefficient. The heavy-electron behavior is possibly associated with the localized Ni$^{3+}$ states in the inner layer of the (NdNiO$_3$)$_3$ block. The different physical properties between La$_4$Ni$_3$O$_{10}$ and Nd$_4$Ni$_3$O$_{10}$ highlight the role of the tolerance factor in controlling the electronic state as well as the lattice distortion.

At least phenomenologically, the MMT of Nd$_4$Ni$_3$O$_{10-\delta}$ bears similarities to those of the Fe-based~\cite{CaFe2As2.cxh} and Ti-based~\cite{2221Ti.neutron,Ba1221.cxh} pnictides. The latter are associated with a density-wave instability, and suppression of the density-wave ordering gives rise to superconductivity~\cite{CaFe2As2.cxh,BaNaTi2Sb2O,BaTi2SbBiO}. External hydrostatic pressure tends to quench the MMT at a finite temperature in Nd$_4$Ni$_3$O$_{10-\delta}$, thus a quantum critical point is unlikely, and superconductivity was not observed down to 1.6 K. Nevertheless, superconductivity could be hopefully realized in the nickelates through a suitable chemical substitution in the future.

\
\begin{acknowledgments}
We thank Yi-feng Yang for helpful discussions. The work at ZJU was supported by the National Key Research and Development Program of China (2017YFA0303002 and 2016YFA0300202) and the Fundamental Research Funds for the Central Universities of China. The work at IOPCAS was supported by the National Key Research and Development Program of China (2018YFA0305700), the National Natural Science Foundation of China (11834016, 11874400), Beijing Natural Science Foundation (Z190008), the Key Research Program of Frontier Sciences of the Chinese Academy of Sciences (QYZDB-SSW-SLH013), and the CAS Interdisciplinary Innovation Team.
\end{acknowledgments}


%

\end{document}